\documentclass[prb,aps,twocolumn,showpacs,superscriptaddress,amsmath,amssymb]{revtex4-1}
\usepackage{amssymb}
\usepackage{amsmath}
\usepackage[pdftex]{graphicx}

\begin{document}

\title{Correlation between Fermi surface transformations and superconductivity in the electron-doped high-$T_c$ superconductor Nd$_{2-x}$Ce$_x$CuO$_4$}

\author{T. Helm}
\altaffiliation{Present address: Material Science Division, Lawrence Berkeley National Laboratory, 94720 Berkeley, CA, USA}

\affiliation{Walther-Mei{\ss}ner-Institut, Bayerische Akademie der Wissenschaften, D-85748 Garching, Germany}

\author{M. V. Kartsovnik}
\email{Mark.Kartsovnik@wmi.badw-muenchen.de}
\affiliation{Walther-Mei{\ss}ner-Institut, Bayerische Akademie der Wissenschaften, D-85748 Garching, Germany}

\author{C. Proust}
\affiliation{Laboratoire National des Champs Magn\'{e}tiques Intenses (LNCMI-EMFL), CNRS, INSA, UJF, UPS, F-31400 Toulouse, France}

\author{B. Vignolle}
\affiliation{Laboratoire National des Champs Magn\'{e}tiques Intenses (LNCMI-EMFL), CNRS, INSA, UJF, UPS, F-31400 Toulouse, France}

\author{C. Putzke}
\altaffiliation{Present address: University of Bristol, Bristol, BS8 1TL, UK}
\affiliation{Hochfeld-Magnetlabor Dresden (HLD-EMFL), Helmholtz-Zentrum Dresden-Rossendorf, D-01328 Dresden, Germany}

\author{E. Kampert}
\affiliation{Hochfeld-Magnetlabor Dresden (HLD-EMFL), Helmholtz-Zentrum Dresden-Rossendorf, D-01328 Dresden, Germany}

\author{I. Sheikin}
\affiliation{Laboratoire National des Champs Magn\'{e}tiques Intenses, (LNCMI-EMFL), CNRS, UJF, F-38042 Grenoble Cedex 9, France}

\author{E.-S. Choi}
\affiliation{National High Magnetic Field Laboratory and Department of Physics,
Florida State University, Tallahassee, Florida 32310, USA}

\author{J. S. Brooks}
\affiliation{National High Magnetic Field Laboratory and Department of Physics, Florida State University, Tallahassee, Florida 32310, USA}

\author{N. Bittner}
\altaffiliation{Present address: Max-Planck-Institut f\"{u}r Festk\"{o}rperforschung, D-70569 Stuttgart, Germany}
\affiliation{Walther-Mei{\ss}ner-Institut, Bayerische Akademie der Wissenschaften, D-85748 Garching, Germany}

\author{W. Biberacher}
\affiliation{Walther-Mei{\ss}ner-Institut, Bayerische Akademie der Wissenschaften, D-85748 Garching, Germany}

\author{A. Erb}
\affiliation{Walther-Mei{\ss}ner-Institut, Bayerische Akademie der Wissenschaften, D-85748 Garching, Germany}
\affiliation{Physik-Department, Technische Universit\"{a}t M\"{u}nchen, D-85748 Garching, Germany}

\author{J. Wosnitza}
\affiliation{Hochfeld-Magnetlabor Dresden, Helmholtz-Zentrum Dresden-Rossendorf, D-01328 Dresden, Germany}

\author{R. Gross}
\email{Rudolf.Gross@wmi.badw-muenchen.de}
\affiliation{Walther-Mei{\ss}ner-Institut, Bayerische Akademie der Wissenschaften, D-85748 Garching, Germany}
\affiliation{Physik-Department, Technische Universit\"{a}t M\"{u}nchen, D-85748 Garching, Germany}

\begin{abstract}
Two critical points have been revealed in the normal-state phase diagram of the electron-doped cuprate superconductor Nd$_{2-x}$Ce$_x$CuO$_4$ by exploring the Fermi surface properties of high quality single crystals by high-field magnetotransport. First, the quantitative analysis of the Shubnikov-de Haas effect shows that the weak superlattice potential responsible for the Fermi surface reconstruction in the overdoped regime extrapolates to zero at the doping level $x_c = 0.175$ corresponding to the onset of superconductivity. Second, the high-field Hall coefficient exhibits a sharp drop right below optimal doping $x_{\mathrm{opt}} = 0.145$ where the superconducting transition temperature is maximum. This drop is most likely caused by the onset of long-range antiferromagnetic ordering. Thus, the superconducting dome appears to be pinned by two critical points to  the normal state phase diagram.
\end{abstract}

\date{\today}
\pacs{74.72.Ek, 71.18.+y, 72.15.Gd}

\maketitle
\section{Introduction}

In order to clarify the mechanism responsible for high-temperature superconductivity in the superconducting (SC) cuprates profound knowledge on the exact nature of the underlying ``normal'', i.e., non-superconducting state is mandatory. This long-standing issue, however, remains controversial. Even for the relatively simple case of the electron-doped cuprates $Ln_{2-x}$Ce$_x$CuO$_4$ ($Ln=$\,Nd, Pr, Sm, La), where the SC state emerges in direct neighbourhood of a state with commensurate antiferromagnetic (AF) order, it is not clear whether the two states coexist and, if yes, to which extent. \cite{armi10} For example, a number of neutron scattering studies have been reported, providing arguments both for \cite{uefu02,fuji08,fuji04,kang05} and against \cite{moto07} the coexistence. Angle-resolved photoemission spectroscopy (ARPES) reveals a reconstruction of the Fermi surface by a long-range commensurate, $\mathbf{Q}=(\pi/a,\pi/a)$, superlattice potential $V_{\mathbf{Q}}$ in the underdoped regime, which possibly survives up to the optimal SC doping level $x_{\mathrm{opt}}\approx 0.15$. \cite{armi02,mats07,sant11,park07}
Magnetotransport studies go even further, indicating the presence of two types of charge carriers \cite{daga04,lin05,balc03,li07,li07a,li07b,lamb08a,helm09,helm10,kart11} and, hence, a reconstructed Fermi surface even in the overdoped region of the phase diagram. Based on the normal-state Hall and Seebeck effects in $Ln=\mathrm{Pr}$ thin films a quantum critical point (QCP) associated with the Fermi surface reconstruction was proposed to lie under the SC dome in the overdoped range of the phase diagram. \cite{daga04,li07b} Finally, studies of the power-law temperature dependence of the resistivity in $Ln=\mathrm{La}$ films \cite{jin11,*butc12} have suggested a QCP at exactly the SC critical doping level $x_c$ on the overdoped side of the phase diagram.

For cuprate superconductors sample quality or surface effects are often invoked to explain apparently contradictory results. Indeed, crystalline quality and doping homogeneity may be serious issues for thin films or for large crystal arrays required, e.g., for neutron scattering experiments. Optical and ARPES techniques are sensitive to surface properties. In this respect, techniques based on magnetic quantum oscillations have obvious advantages: they probe bulk properties and can be performed on small single crystals, whose high quality is already ensured by the very existence of quantum oscillations.

The first experiment on magnetoresistance quantum oscillations (Shubnikov-de Haas, SdH effect) in Nd$_{2-x}$Ce$_x$CuO$_4$ (NCCO) crystals \cite{helm09} apparently corroborated the existence of a
QCP hidden under the SC dome on the overdoped side, i.e. between $x_{\mathrm{opt}}$ and $x_c$. However, subsequent more elaborate SdH experiments \cite{helm10,kart11} have further extended the range in which the Fermi surface stays reconstructed up to at least $x=0.17$, the highest (though still SC) doping level attainable in bulk NCCO crystals. Thus, while the existence of one or even more critical points associated with a Fermi surface reconstruction in the electron-doped cuprates is generally accepted, \cite{armi10,das07,*das08,sach10,*sach09,chub13} the question about their location with respect to the SC part of the phase diagram and their possible relation to the SC pairing remains open.

Here, we report systematic high-field magnetotransport studies on high-quality NCCO crystals, which allow us to precisely locate two critical doping levels in the normal-state phase diagram of this material and correlate them with the position of the SC dome. First, by performing a quantitative analysis of SdH oscillations observed in the magnetic-breakdown (MB) regime on overdoped samples, we evaluate the small MB gap $\Delta_{\mathrm{MB}} \simeq V_{\mathbf{Q}}$, separating the hole and electron pockets of the reconstructed Fermi surface, as a function of $x$. The $x$-dependence of the gap is found to mimic that of the SC critical temperature $T_c(x)$, both extrapolating to zero at the same characteristic doping level $x_c \approx 0.175$. Second, we present high-field Hall resistance measurements which, in combination with the SdH data, reveal a large energy gap emerging in the system right below the optimal doping level $x_{\mathrm{opt}}=0.145$. Although the present data alone cannot give a direct key to the microscopic origin of the detected Fermi surface transformations, they provide a strong evidence of the importance of these transformations for superconductivity.

\section{Experimental procedures}
\begin{figure}[h]
\includegraphics[width=1.0\columnwidth]{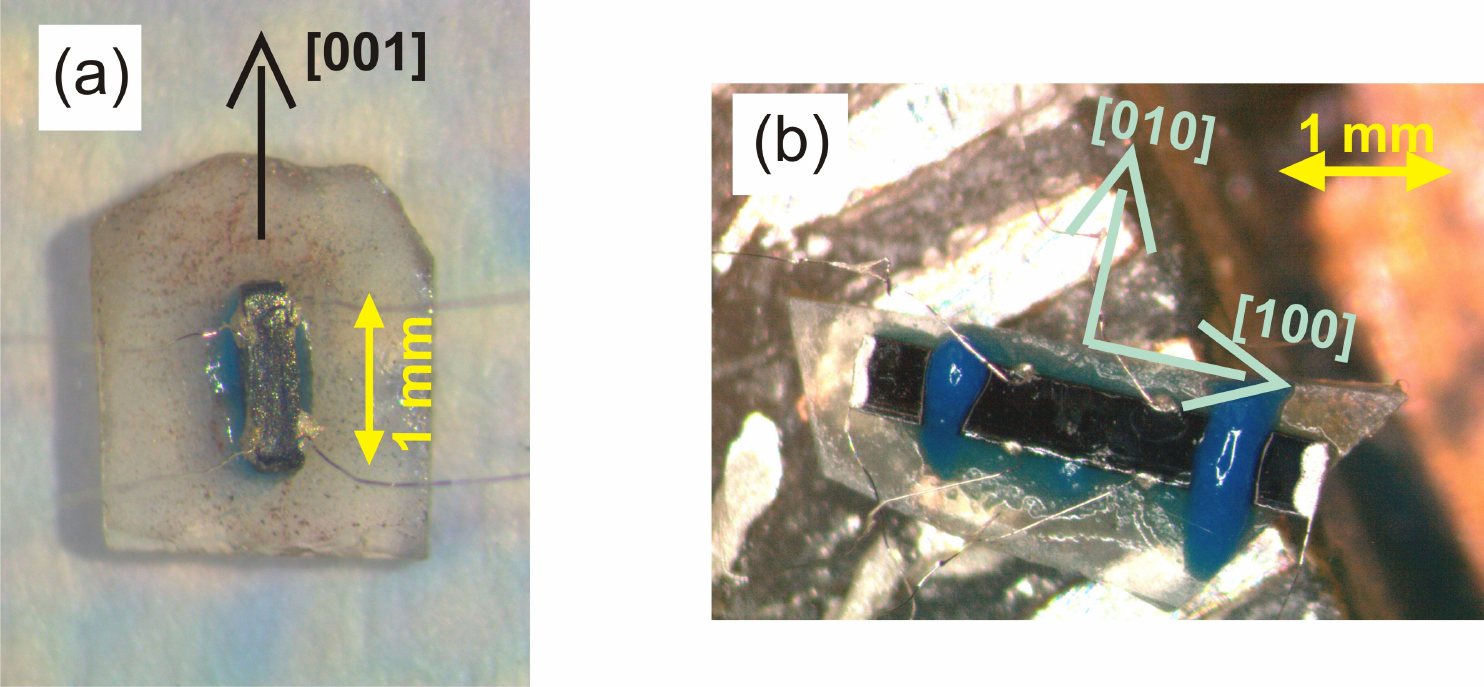}
\caption{(Color online) NCCO single crystals prepared for measuring (a) the 4-point interlayer resistance and (b) the in-plane Hall resistance. In both cases thin Pt wires are glued by silver-based epoxy to the samples surface and the samples are fixed to a sapphire substrate by Stycast 2850FT (blue). }
\label{Helm_Supplement_Figure1}
\end{figure}

The NCCO crystals with Ce concentrations $0.13 \leq x \leq 0.17$ were grown using the travelling solvent floating zone technique, annealed, and characterized as described in Ref. \citenum{lamb10}. For each $x$, several single crystals showing the best residual resistance ratios and narrow SC transitions were selected. The electrical leads for magnetotransport experiments were made using 10\,- or 20\,-$\mu$m-diameter annealed platinum wires glued to the crystals using the Epotek H20E silver-based conductive epoxy. The typical contact resistance was $\sim 1-10\,\Omega$. Different sample shapes were chosen in order to optimize the geometrical configuration for the quantum oscillation and Hall effect measurements, respectively. Quantum oscillations of the interlayer resistance were measured on samples having a small, $\sim 0.1$\,mm$^2$, cross section in the plane of CuO$_2$ layers and a length of $1-2$\,mm along the $c$-axis, see Fig.~\ref{Helm_Supplement_Figure1}(a). In the Hall effect studies the current was applied along the CuO$_2$ layers. The samples were prepared in the shape of a thin plate with a thickness of $0.05$\,mm along the $c$-axis. The width and length of the plates were $0.2-0.3$\,mm and $2-5$\,mm in the $a$- and $b$-direction, respectively [see~Fig.~\ref{Helm_Supplement_Figure1}(b)]. Particular care was taken to achieve low-resistance electrical contacts between the current leads and the sample edges, ensuring a homogeneous current flow.

 All the experiments were done in magnetic fields perpendicular to the CuO$_2$ layers. Most of the results were obtained in pulsed magnetic fields. Additionally, measurements in steady fields at precisely controlled temperatures were done for evaluating effective cyclotron masses. In the Hall effect experiments magnetic-field sweeps of opposite polarities were always made in order to eliminate the even magnetoresistance component.

\section{SdH effect in the MB regime}
\subsection{Experimental results and analysis}
\begin{figure}[tb]
\includegraphics[width=1.0\columnwidth]{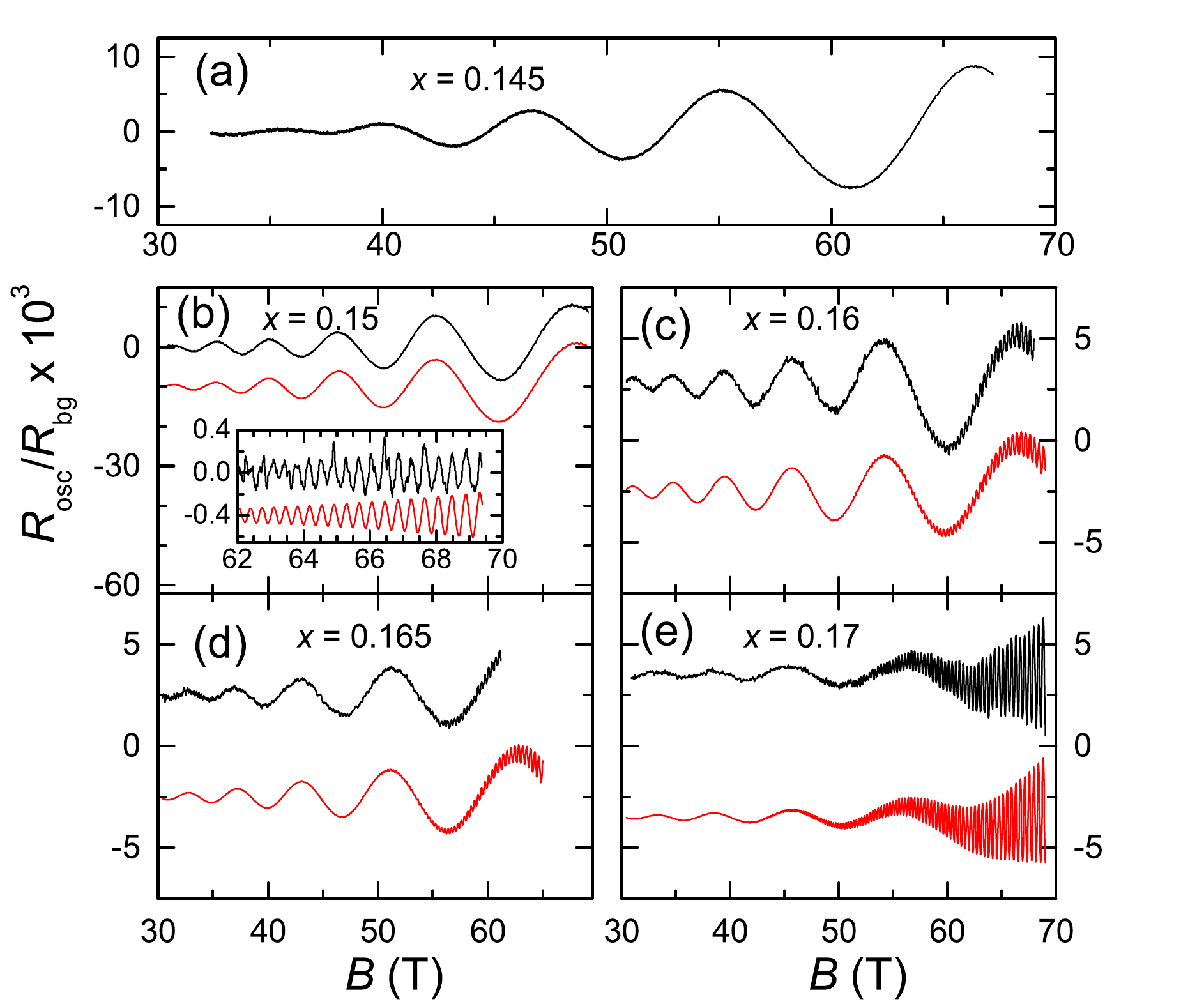}
\caption{(Color online) Black lines: SdH oscillations obtained on NCCO crystals with different doping levels $x$ at $T\approx 2.5$\,K. The data are normalized to the nonoscillating field-dependent background resistance $R_{\mathrm{bg}}$. Red lines: fits to the experimental data made as described in the text. The fitting parameters are given in Table\,\ref{Table}. The curves are vertically shifted for clarity. Inset in panel (b): enlarged view of the fast MB oscillations for $x=0.15$ after subtracting the slowly oscillating component.}
\label{SdH}
\end{figure}

Figure\,\ref{SdH} shows examples of SdH oscillations in the interlayer resistance of optimally doped and overdoped NCCO at $T \approx 2.5$\,K. For $x\geq 0.15$, the oscillations contain two characteristic frequencies: $F_{\alpha} \simeq 250-300$\,T and $F_{\beta} \simeq 11$\,kT. The slow $\alpha$ oscillations are associated with orbits on the small hole pockets of the reconstructed Fermi surface. \cite{helm09} The fast $\beta$ oscillations reveal a cyclotron orbit, which is geometrically equivalent to that on the large unreconstructed Fermi surface and arises from the MB effect. \cite{helm10,kart11,chak11} The fast oscillations are dominant at high fields for $x=0.17$ and rapidly diminish at decreasing doping. For $x=0.15$, they are about  100-times weaker than the slow oscillations at the same field strength. At optimal doping, $x_{\mathrm{opt}} =0.145$, the $\alpha$ oscillations are $\sim 20\%$ weaker than at $x=0.15$, whereas the $\beta$ oscillations are no
longer resolvable above the noise level, $\lesssim 10^{-4}$ of the total resistance.

On the qualitative level the observed behavior is easily understood as a result of an enhancement of the superlattice potential $V_{\mathbf{Q}}$, hence, of the MB gap with decreasing $x$. Moreover, due to the very good signal-to-noise ratio of our measurements, the data can be analyzed quantitatively, allowing us to estimate the MB gap as a function of $x$.
To this end, we have applied the standard Lifshitz-Kosevich (LK) formalism, \cite{abri88b,shoe84} additionally taking into account the MB effect.

Since only the fundamental harmonics $F_{\alpha}$ and $F_{\beta}$ were observed, the analysis can be restricted to the first harmonic term of the LK expansion:
\begin{equation}
\frac{\sigma_{\mathrm{osc},j}}{\sigma_{\mathrm{bg},j}} = A_j \cos\left(2\pi\frac{F_j}{B}+\gamma_j\right) \,,
\label{LK}
\end{equation}
where $\sigma_{\mathrm{osc}}$ and $\sigma_{\mathrm{bg}}$ are the oscillating and non-oscillating (background) components of the interlayer conductivity, respectively. The index $j=\alpha,\beta$ is used to label the slow (classical) and fast (MB) oscillations, respectively.

The oscillation amplitude is considered in the form:
\begin{equation}
A_j(B,T) = A_0\mu_{j} B^{1/2} R_{T,j}(B,T)R_{D,j}(B)R_{\mathrm{MB},j}(B) \,,
\label{A}
\end{equation}
where $A_0$ is a field- and temperature-independent factor, $\mu_j=m_{c,j}/m_0$ is the effective cyclotron mass in units of the free electron mass;
the LK temperature damping factor is: \cite{lifs55}
\begin{equation}
R_{T,j} = \frac{K\mu_{j}T/B}{\sinh(K\mu_jT/B)}
\label{RT}
\end{equation}
with $K=2\pi^2k_{\mathrm{B}}m_0/\hbar e \approx 14.69$\,T/K, and the Dingle (scattering) damping factor is:
\begin{equation}
R_{D,j} = \exp(-K\mu_jT_D/B)\, .
\label{RD}
\end{equation}
Here, $T_D$ is the Dingle temperature characterising the Landau-level broadening caused by a finite quasiparticle lifetime \cite{shoe84,ding52,bych61a} (for the relevant low temperatures it is mainly determined by crystal imperfections). In addition to the standard damping factors, Eq. (\ref{A}), contains the factor $R_{\mathrm{MB}}$ which takes into account the MB effect. It is expressed as: \cite{fali66} $R_{\mathrm{MB},j}=(\imath \upsilon)^{l_{1,j}}\xi^{l_{2,j}}$, where $\upsilon$ and $\xi$ are, respectively, the probability amplitudes for tunneling through and Bragg reflection at a MB junction. These amplitudes are determined by the MB field $B_0$ as $\upsilon = \exp(-B_0/2B)$ and $\xi= \left[1-\exp(-B_0/B)\right]^{1/2}$. The exponents $l_{1,j}$ and $l_{2,j}$ are, respectively, the numbers of the MB junctions through which the charge carriers should tunnel and at which they should be reflected in order to complete the $j$-th orbit. In our case of the Fermi surface reconstructed by a $(\pi/a,\pi/a)$ potential, the small classical $\alpha$ orbit involves two reflections at MB junctions. \cite{kart11} Therefore, $l_{1,\alpha}=0$, $l_{2,\alpha}=2$ and the corresponding reduction factor is
\begin{equation}
R_{\mathrm{MB},\alpha} = \left[1-\exp(-B_0/B)\right]\,.
\label{MBa}
\end{equation}
The large $\beta$ orbit involves tunneling through eight MB junctions, \cite{kart11} so that $l_{1,\beta}=8$, $l_{2,\beta}=0$ and
\begin{equation}
R_{\mathrm{MB},\beta}=\exp(-4B_0/B)\,.
\label{MBb}
\end{equation}
It is immediately evident that an increase of $B_0$ leads to an increase of $R_{\mathrm{MB},\alpha}$ and, in turn, to a decrease of $R_{\mathrm{MB},\beta}$.

For fitting the experimental data, the oscillation frequencies and phase factors entering Eq.\,(\ref{LK}) were determined directly from the positions of the minima and maxima of the measured SdH oscillations.

As a next step, the cyclotron masses $\mu_j$ were evaluated using the $T$-dependence of the oscillation amplitudes determined by Eq.\,(\ref{RT}). The mass corresponding to the slow $\alpha$ oscillations was obtained both from steady- and from pulsed-field measurements. The advantage of the steady-field measurements is that the temperature can directly be determined with high accuracy. In pulsed fields there are issues related to overheating due to eddy currents, fast magnetization changes, and relatively high (up to 10\,mA) measurement currents. Therefore, special care was taken for good thermalization of the samples in liquid helium during the field pulses. The data for the analysis was taken from the relatively slow, $\sim 0.15$\,s, decaying part of the pulse. The temperature was determined from the resistive superconducting (SC) transitions by comparing them to the transitions recorded in steady fields and at low, $\sim 0.1$\,mA, measurement currents.

\begin{figure}[tb]
\includegraphics[width=1.0\columnwidth]{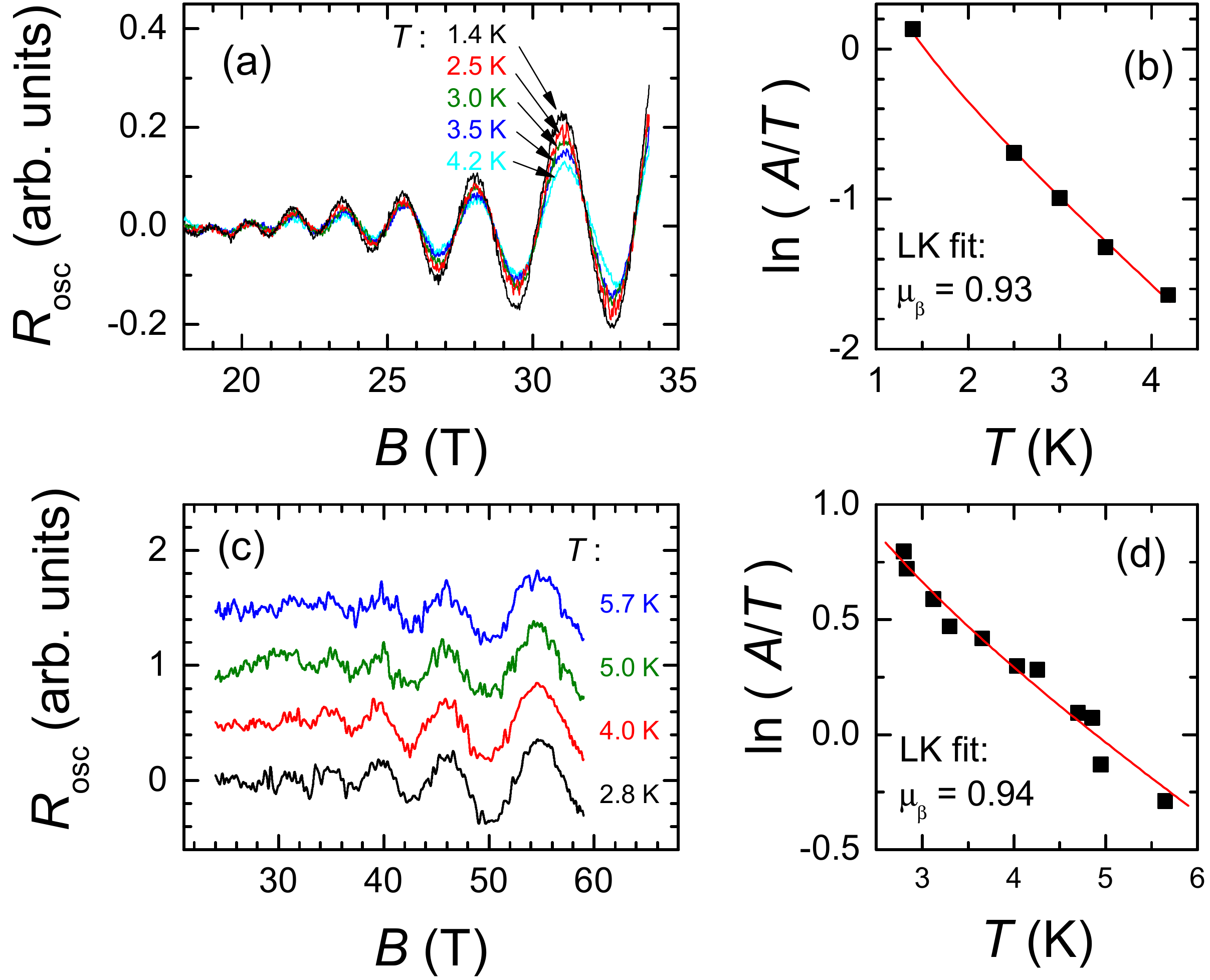}
\caption{(Color online) (a) Slow SdH oscillations of the interlayer resistance of a NCCO single crystal with $x=0.16$ recorded in steady magnetic fields at different temperatures. (b) The mass plot of the FFT  amplitudes $A$ of the data in (a) and its fit to Eq.~(\ref{RT}) yielding a normalized cyclotron mass of 0.93. Panels (c) and (d) present the same plots as in (a) and (b) but for the data obtained in pulsed fields.}
\label{Helm_Supplement_Figure2}
\end{figure}

An example of the
$\alpha$ oscillations observed on a $x=0.16$ sample at different temperatures, both in steady and in pulsed fields is shown in Fig.~\ref{Helm_Supplement_Figure2}(a) and (c).
Panels (b) and (d) of Fig.\,\ref{Helm_Supplement_Figure2} show the temperature dependence ("mass plot") of the fast Fourier transform (FFT) amplitudes of the oscillations, demonstrating a very good agreement between both data sets. The same good agreement was obtained for $x=0.145, 0.15$, and 0.165. For $x=0.17$, a determination of the cyclotron mass was not possible due to a very small oscillation amplitude and a flat $T$-dependence. For the further analysis the value $\mu_{\alpha}^{0.17}=0.88$ extrapolated from lower doping was taken.

\begin{figure}[tb]
\includegraphics[width=1.0\columnwidth]{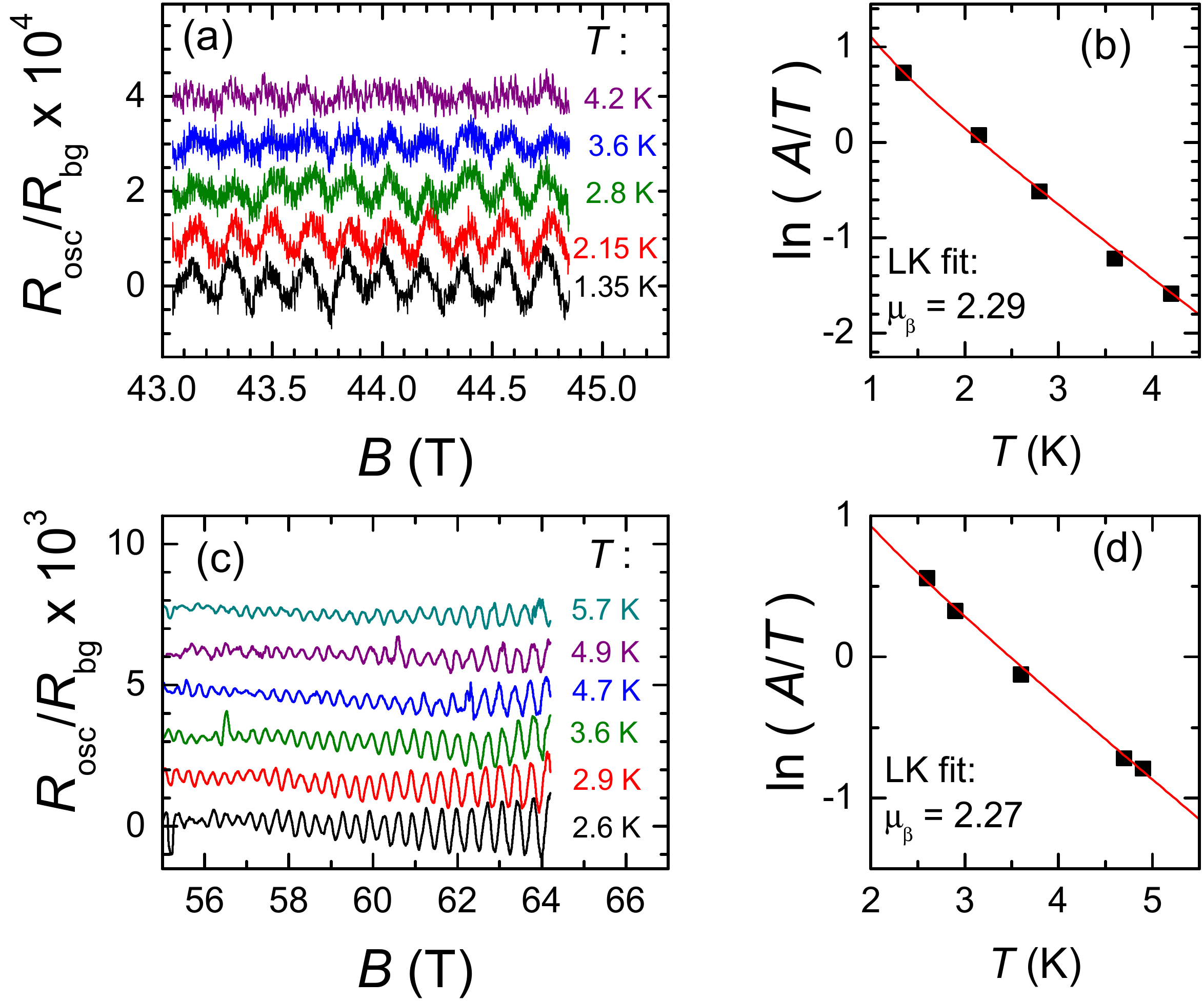}
\caption{(Color online) (a) Fast SdH oscillations in a $x=0.17$ sample in steady magnetic fields at different temperatures. (b) the corresponding mass plot of the oscillation amplitude, yielding a normalized cyclotron mass of $\approx 2.3$. The panels (c) and (d) show the same plots as in (a),(b) but for the data obtained in pulsed fields. }
\label{Helm_Supplement_Figure3}
\end{figure}
Although the MB $\beta$ oscillations are generally significantly weaker than the $\alpha$ oscillations and vanish faster with decreasing the magnetic field, we succeeded in detecting them on the strongly overdoped, $x=0.17$ crystal not only in pulsed fields but also in steady fields of up to 45\,T. The oscillation patterns obtained at several temperatures in a field window of 43 to 45\,T and the corresponding mass plot yielding $\mu_{\beta}\approx 2.3$ are shown in Fig.~\ref{Helm_Supplement_Figure3}(a),(b). Again, as in the case of the slow oscillations, this data is in very good agreement with the pulsed field data shown in Fig.~\ref{Helm_Supplement_Figure3}(c) and (d). This provides a solid justification for the validity of the pulsed field data in the cyclotron mass analysis at lower doping levels.

Before proceeding to fitting the experimentally obtained field dependence of the oscillations by Eqs.\,(\ref{LK}) and (\ref{A}) one has to express the resistance oscillations in terms of conductivity and determine the relative contributions of the channels responsible for the $\alpha$ and $\beta$ oscillations to the background conductivity, as it is explained in the Appendix. Since the MB $\beta$ oscillations involve all the carriers on the Fermi surface, the corresponding background is simply the total conductivity: $\sigma_{\mathrm{bg},\beta} = \sigma_{\mathrm{bg}} \propto 1/R_{\mathrm{bg}}$ and, as discussed in the Appendix, quadratically decreases at increasing field. The contribution of the channel associated with the $\alpha$ oscillations can be estimated as $\approx 0.4 \sigma_{\mathrm{bg}}(0)$ at zero field and is approximately constant at high magnetic fields (see Appendix).

 Knowing the relative contributions of the $\alpha$ and $\beta$ channels to the interlayer conductivity, the effective cyclotron masses, as well as the frequencies and phase factors of the oscillations, we fitted the oscillatory resistance traces $R_{\mathrm{osc}}(B)/R_{\mathrm{bg}}(B)$ experimentally obtained at a fixed temperature, using $A_0$, $T_D$, and $B_0$ in Eqs.~(\ref{A}) and (\ref{RD})-(\ref{MBb}) as free parameters. \cite{comment_Dingle} Special care was taken to reproduce not only the field dependence but also the relative amplitudes of the $\alpha$ and $\beta$ oscillations.

 Figure \ref{SdH} shows the fitting results (red curves) in comparison with the experimentally observed oscillations (black curves) at different doping levels. The main oscillation parameters obtained from the fits are presented in Table\,\ref{Table}. No fitting was done for $x=0.145$, since no fast $\beta$ oscillations have been resolved for this doping level. For the other 4 doping levels the fits nicely reproduce the relative amplitudes as well as the field dependence of both $\alpha$ and $\beta$ oscillations.

\begin{table}[tb]
\caption{\label{Table}
Oscillation frequencies, effective cyclotron masses (in units of the free electron mass), Dingle
temperatures, and MB fields obtained for NCCO crystals with different Ce concentrations $x$ from fitting the experimental data on the SdH oscillations.}
\begin{ruledtabular}
\begin{tabular}{ccccccc}
x & $F_{\alpha}$\,[T] & $F_{\beta}$\,[T] & $\mu_{\alpha}$ & $\mu_{\beta}$ & $T_D$\,[K] & $B_0$\,[T]\\
\hline
0.17 & 246	 & 10935		& $0.88\footnote{extrapolated from lower doping}$ 	& 2.3		&   13.8      	&   1.0 \\
0.165 & 275       &   11030         &        0.90     								& 2.5  	&   16.6       &   1.5 \\
0.16 &  290        &   11155         &        0.92     								& 2.7  	&   13.5    &   3.0   \\
0.15 &  292        &   11250         &        1.05    								& 3.0  	&   13.5    &   12.5 \\
0.145 & 295       &     --          	&         1.30     								&  --    	&    12.3    &   -- \\
\end{tabular}
\end{ruledtabular}
\end{table}

\subsection{Doping dependence of the oscillation parameters: evidence of two critical points}
\begin{figure}[tb]
\includegraphics[width=1.0\columnwidth]{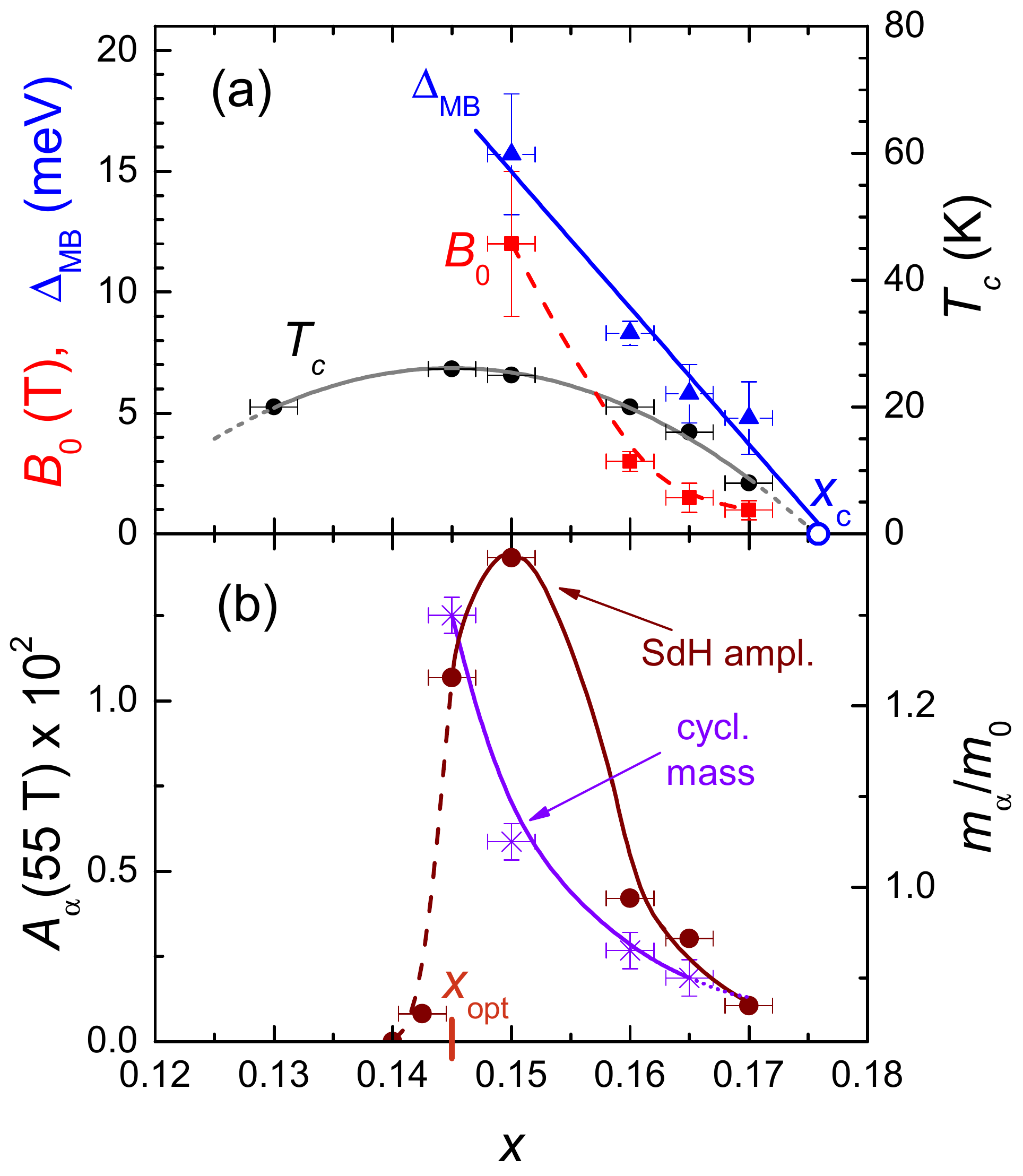}
\caption{(Color online) (a) MB field $B_0$ (squares), the corresponding energy gap $\Delta_{\mathrm{MB}}$ (triangles), and the SC critical temperature $T_c$ plotted as a function of Ce concentration $x$. The straight line is a linear fit to the $\Delta_{\mathrm{MB}}(x)$ dependence, extrapolating to zero at $x_c\approx 0.175$. The other lines are guides to the eye.
(b) $x$-dependence of the amplitude $A_{\alpha}$ of the slow oscillations measured at $B=55$\,T, $T=2.5$\,K (circles) and the relevant cyclotron mass $m_{\alpha}$ normalized to the free electron mass $m_0$ (stars). The lines are guides to the eye.}
\label{properties}
\end{figure}

The values of the MB field $B_0(x)$ obtained from fitting are also plotted in Fig.\,\ref{properties}(a) by squares.
From this data the energy gap $\Delta_{\mathrm{MB}}$ between different parts of the reconstructed Fermi surface can be estimated according to Blount's criterion \cite{shoe84} $\Delta_{\mathrm{MB}} \approx \left(\hbar e B_0  \varepsilon_{\mathrm{F}}/m_{\beta}\right)^{1/2}$. Here, $e$ is the elementary charge and the Fermi energy is \cite{mark05} $\varepsilon_{\mathrm{F}}\approx 0.5$\,eV. The $\Delta_{\mathrm{MB}}$ values (blue triangles) are plotted in Fig.\,\ref{properties}(a) as a function of $x$ along with the SC critical temperature $T_c(x)$ (black circles, right-hand scale). We see that the MB gap is small (meV range) and decreases approximately linearly with increasing $x$ in the overdoped regime. Most importantly, $\Delta_{\mathrm{MB}}(x)$ extrapolates to zero at the same characteristic doping level $x_c = 0.175$, at which $T_c$ is believed to vanish. \cite{lamb10,taka89}

Figure\,\ref{properties}(b) shows the doping dependence of the amplitude $A_{\alpha}$ of the $\alpha$ oscillations normalized to the non-oscillating resistance background together with the corresponding effective cyclotron mass $m_{\alpha}$. Obviously, $m_{\alpha}$ increases rapidly on moving from the strongly overdoped regime towards optimal doping. This change cannot simply be explained by the experimentally observed \cite{kart11} weak $x$-dependence of the area of the $\alpha$ orbit. It is most likely a mass renormalization effect due to enhanced electron correlations in the vicinity of a metal-insulator transition.
In fact, it resembles the behavior observed recently in hole-doped cuprate \cite{seba10a} and iron-pnictide \cite{walm13,shib14} superconductors near a quantum critical point.

The $x$-dependence of $A_{\alpha}$ in the overdoped regime is governed by that of the MB gap: it rapidly grows upon going from $x=0.17$ to $0.15$. At optimal doping it is slightly smaller than at $x=0.15$, which is consistent with the considerable, $\sim 20\%$, increase of the cyclotron mass and consequent reduction of the temperature and Dingle factors in the expression for the oscillation amplitude. \cite{shoe84} A further decrease of $x$ leads to a dramatic suppression of $A_{\alpha}$. When the doping level is decreased by just $\Delta x \approx 0.03\%$ below $x_{\mathrm{opt}}$, the amplitude drops by more than an order of magnitude, becoming too small for a quantitative analysis. At present, it cannot even be ruled out that the weak oscillations remaining at $x=0.142$ are caused by a minor optimally doped sample fraction due to an unavoidable small inhomogeneity of the Ce distribution. Assuming for a moment that the oscillations are, however, inherent to a perfectly homogeneous $x=0.142$ sample, we estimate that the cyclotron mass should increase by almost a factor of 2 at decreasing $x$ by $0.03\%$, in order to account for the observed reduction of the amplitude. Such a steep rise would be a strong argument in favor of the mass divergence near the optimal doping level.

An alternative mechanism for the observed suppression might involve an abrupt change in the electronic spectrum or in scattering processes. We note that a trivial scenario associated with a poor crystal quality is highly unlikely in our case. From the crystal growth point of view, \cite{lamb08a,erb14} the sample quality should not vary considerably with $x$ around $x_{\mathrm{opt}}$. Consistently, the Dingle temperatures, sensitive to crystal imperfections, \cite{shoe84} obtained from our fits in Fig.\,\ref{SdH} are close to each other, $T_D = 13 \pm 1$\,K, for crystals with $x=0.145$ to 0.16 (see Table\,\ref{Table}). This suggests that also for slightly underdoped samples, $0.14 \leq x < 0.145$, quality is unlikely a critical issue. Hence, the reason for the suppression of the quantum oscillations must lie in an intrinsic significant change in the electronic system.

\section{Low-temperature, high-field Hall effect}
\begin{figure}[t]
\includegraphics[width=0.9\columnwidth]{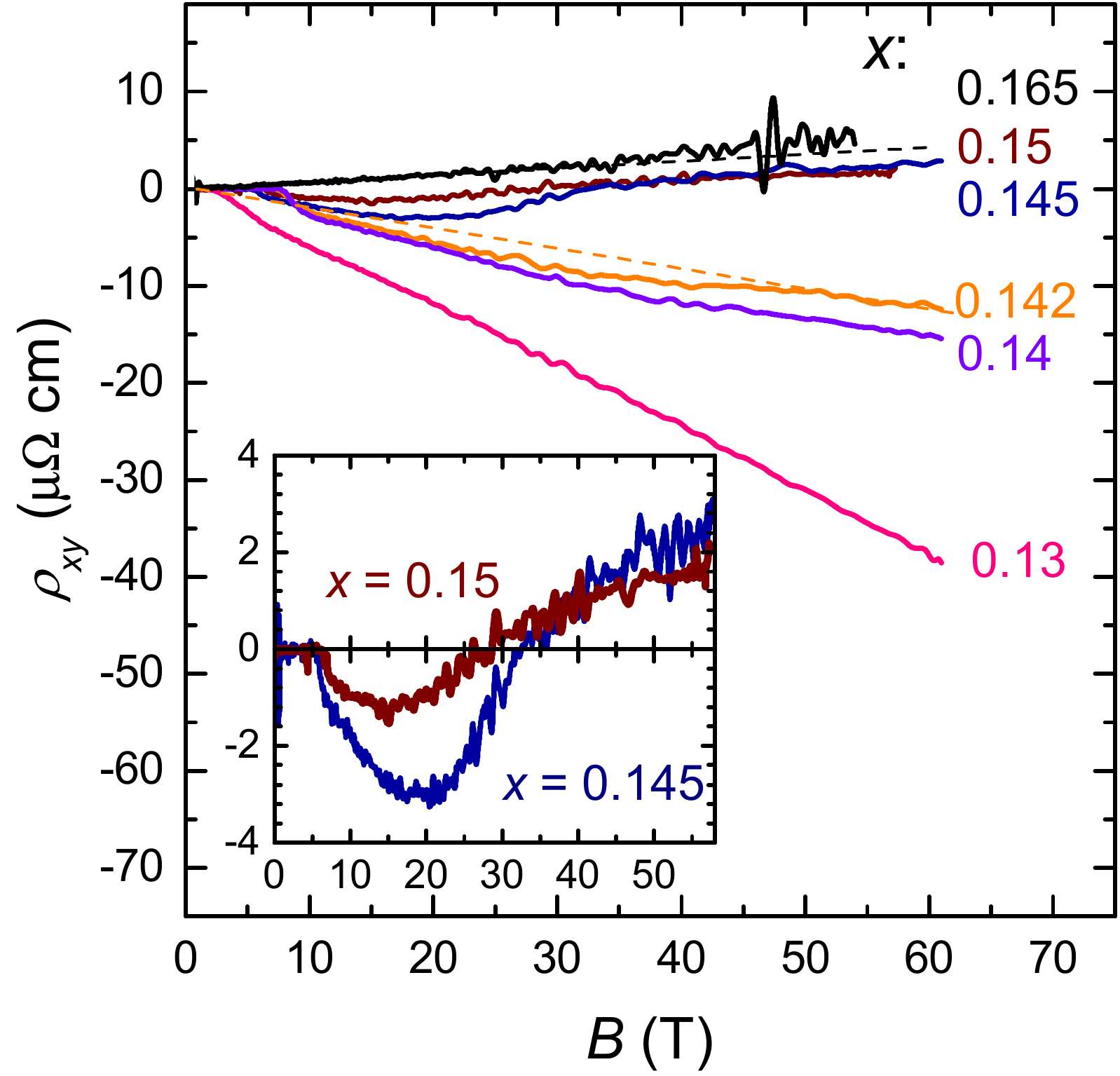}
\caption{(Color online) Field-dependent Hall resistivity at $T=2$\,K, for different $x$. Inset: enlarged view on the data for $x=0.145$ and 0.15.}
\label{Hall_B}
\end{figure}

To gain further insight in this change, we have studied the high-field Hall effect in NCCO crystals with different Ce concentrations, focusing on the regime around  $x_{\mathrm{opt}}$. Figure\,\ref{Hall_B} shows examples of the field-dependent Hall resistivity $\rho_{xy}(B)$ measured at $T \approx 2$\,K. At this low temperature various complications associated with inelastic scattering and thermal fluctuations \cite{kont08} can be neglected. Outside the very narrow range around $x_{\mathrm{opt}}$, our data is in good agreement with previous studies on NCCO single crystals \cite{wang05} and on thin films of the sister compound Pr$_{2-x}$Ce$_x$CuO$_4$ (PCCO). \cite{daga04,li07a, char10} Note that the small positive Hall resistivity measured for the overdoped ($x=0.165$) sample, indicating a single large holelike orbit, is fully consistent with a  multiply-connected reconstructed Fermi surface, if one takes into account the very low MB field $B_0 \approx 1.5$\,T determined for this doping.

Important new features have been detected in the close vicinity of $x_{\mathrm{opt}}$. A spectacular manifestation of the MB effect is the nonmonotonic $\rho_{xy}(B)$ dependence obtained for $x=0.15$ and $0.145$ (see inset in Fig.\,\ref{Hall_B}). Here the MB field is moderately strong, $B_0 \approx 12$\,T for $x=0.15$. At $B<B_0$ the MB probability is low and the behavior is qualitatively described by the classical two-band model neglecting the MB effect. \cite{lin05} The normal-state Hall conductivity is determined by competing contributions from electron- and hole-like orbits on the reconstructed Fermi surface, resulting in a small negative $\rho_{xy}(B)$. At $B\gtrsim B_0$ the MB probability becomes significant. Therefore $\rho_{xy}(B)$ turns up, crosses zero, and eventually assumes a linear positive slope in the strong MB regime, where the large holelike orbit dominates like in the case of strongly overdoped samples. A comparison of the two curves shown in the inset of Fig.\,\ref{Hall_B} suggests that at $x_{\mathrm{opt}}$ the MB field is $\sim 5$\,T higher than at $x=0.15$. Interestingly, the nonmonotonic shape of $\rho_{xy}(B)$ clearly correlates with the
anomalous magnetoresistance behavior observed near optimal doping. \cite{helm09,li07a} Thus, the latter anomaly is obviously also associated with the MB effect.

\begin{figure}[t]
\includegraphics[width=0.9\columnwidth]{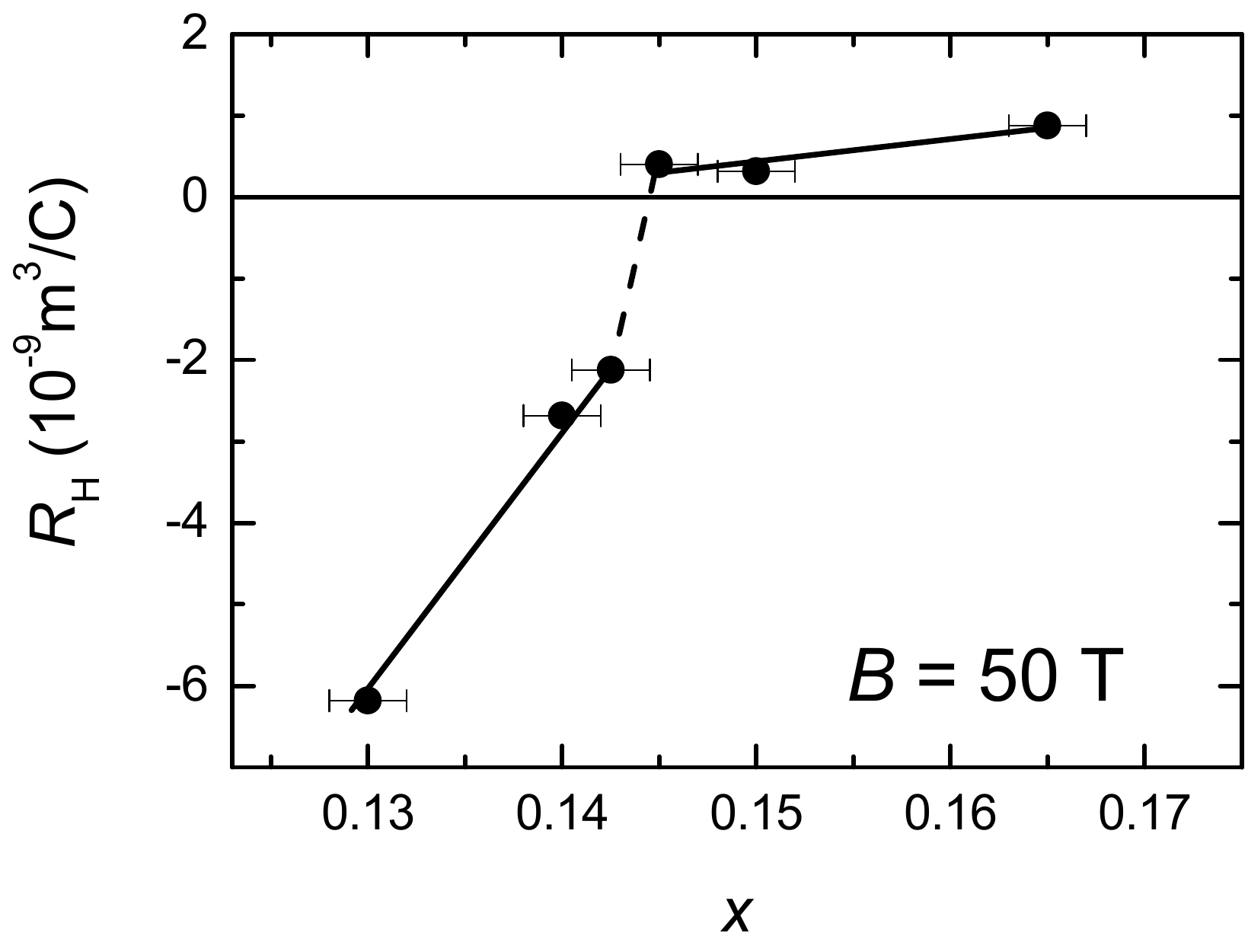}
\caption{Hall coefficient measured at $B=50$\,T and $T=2$\,K as a function of $x$. Lines are guides to the eye.}
\label{Hall_B2}
\end{figure}

The most remarkable result of our Hall effect study is the fact that $\rho_{xy}(B)$ changes dramatically on reducing the doping level below $x_{\mathrm{opt}}$. Already for $x=0.142$, the weak positive signal observed for $x \geq x_{\mathrm{opt}}$ at high fields is replaced by a large negative signal with no sign of saturation at the highest fields. \cite{comment_Hall} This change is especially manifested in a sharp step in the $x$ dependence of the high-field Hall
coefficient $R_{\mathrm{H}} = \rho_{xy}/B$ observed between
$x=0.145$ and 0.142, as shown in Fig.\,\ref{Hall_B2}. The negative linear slope
of $\rho_{xy}(B)$ obtained for $x\leq 0.142$ up to the highest fields
indicates that no MB occurs in the underdoped regime.
This means that the gap between different parts of the reconstructed Fermi surface
sharply increases within the very narrow doping interval right below
$x_{\mathrm{opt}}$.

\section{Discussion}
The data presented above indicate the presence of two critical points in the phase diagram of NCCO whose positions clearly correlate with the position of the SC dome. At first glance this result confirms the theoretical prediction of two topological transformations of the Fermi surface occurring at $x_{\mathrm{opt}}$ and $x_c$, respectively. \cite{das07,*das08} In fact, the situation is less obvious. The calculations \cite{das07,*das08} predict a Lifshitz transition associated with vanishing of the small hole Fermi pockets upon decreasing $x$ below $x_{\mathrm{opt}}$. On the one hand, this might explain the sudden suppression of the SdH oscillations seen in Fig.\,\ref{properties}(b). On the other hand, our SdH data do not reveal any significant decrease of the size of the hole pockets, which should precede the Lifshitz transition, at approaching $x_{\mathrm{opt}}$. Thus, while the Hall data indicates a sharp increase of the MB gap near $x_{\mathrm{opt}}$, this unlikely leads to a complete collapse of the hole pockets.

It is natural to attribute the sharp increase of $\Delta_{\mathrm{MB}}$
to an onset of the static AF order coexisting with superconductivity below $x_{\mathrm{opt}}$.
This is apparently in line with the ARPES data, \cite{armi02,mats07,sant11} implying a Fermi surface reconstruction due to an AF superlattice potential persisting up to $x_{\mathrm{opt}}$. It was argued that a spurious magnetic superstructure signal in SC NCCO might come from minor epitaxial precipitations of paramagnetic (Nd,Ce)$_2$O$_3$ unavoidably present in oxygen-reduced crystals \cite{mang04a} or from remnants of an insufficiently reduced phase. \cite{moto07} However, our transport data, insensitive to insulating precipitations, unambiguously reveal the gap as an inherent feature of the major conducting phase, setting in right below optimal doping.

Turning to the overdoped regime, the $x$-dependence of the small MB gap in Fig.\,\ref{properties}(a) gives strong support to the proposed \cite{das07,*das08} quantum phase transition at the critical SC overdoping $x_c$. Taken together with the recent report on a QCP detected at the same location in La$_{2-x}$Ce$_x$CuO$_4$ thin films, \cite{jin11} it appears to be a general property of electron-doped cuprates. Our SdH results clearly identify this transition as a Fermi surface reconstruction caused by translation symmetry breaking. However, the nature of the relevant ordering is still unclear. While, as argued above, static antiferromagnetism is most likely established right below optimal doping, no convincing evidence for it has been found at $x_{\mathrm{opt}}<x<x_c$. \cite{armi10} Possible alternatives can be a hidden $d$-density-wave order \cite{chak01} or recently discovered charge ordering. \cite{dama15} Another possibility to consider is that the ordered state is induced by a strong magnetic field. \cite{chen08,sach10,*sach09}

On the other hand, the observation of the slow SdH oscillations (and thereby a finite MB gap) may be consistent with a fluctuating AF order reported by several groups, \cite{armi10,moto07,fuji08} provided the corresponding time scale and correlation length are sufficiently large.
A lower-limit estimate for the time over which a charge carrier ``sees" the potential $V_{\mathrm{Q}}$ is obtained from the Dingle temperature $T_D$. For optimally doped samples $T_D \approx 13$\,K, yielding $\tau_D = \hbar/2\pi k_B T_D \approx 0.9\times 10^{-13}$\,s. The corresponding low limit for the correlation length,
$\xi_{\mathrm{min}}\sim \tau_D v_F \simeq 18$\,nm, is $\sim 45$ times the unit cell period, an order of magnitude larger than those reported for the AF \cite{moto07} and charge \cite{dama15} ordering. Therefore the exact origin of the ordered state on the overdoped side of the phase diagram is still an open question.

\section{Conclusion}
The present study reveals the existence of two critical points in the normal-state phase diagram of NCCO. The doping values of these points remarkably correlate with those characterizing the SC dome. On reducing $x$, superconductivity emerges at the same critical doping level, $x_c\approx 0.175$, as the weak superlattice potential $V_{\mathbf{Q}}\sim \Delta_{\mathrm{MB}}$. Both $V_{\mathbf{Q}}$ and $T_c$ grow at decreasing doping towards the optimal value $x_{\mathrm{opt}}$. Thus, while the exact origin of $V_{\mathbf{Q}}$ is still to be determined, it obviously must have a strong impact on the SC pairing. The optimal SC doping coincides with the second critical point where a large energy gap sets in. This can naturally be explained by an intrinsic competition between superconductivity and long-range antiferromagnetism. As argued above, the large energy gap is an inherent feature of the major conducting phase. A highly interesting question is related to the possible microscopic coexistence of antiferromagnetism and superconductivity in high-quality underdoped NCCO crystals. Further studies are required to settle this issue.

\begin{acknowledgments}
This work was supported by the German Research Foundation via grant GR\,1132/15. We acknowledge support of our high-field experiments by HLD-HZDR (Dresden) and LNCMI-CNRS (Toulouse, Grenoble), members of the European Magnetic Field Laboratory. Part of experiments was performed at the NHMFL (Tallahassee), under the support by NSF-DMR 1005293, NSF Cooperative Agreement No. DMR-0654118, the State of Florida, and the U.S. Department of Energy.
\end{acknowledgments}

\begin{appendix}

\section{Relative contributions of the $\alpha$ and $\beta$ conduction channels to the oscillatory conductivity and monotonic background}
%\label{AppA}

Thanks to the very high electronic anisotropy of NCCO (the resistivity anisotropy ratio is \cite{armi10} $\rho_c/\rho_{ab} \gtrsim 10^3$), its interlayer resistance is simply inversely proportional to the interlayer conductivity even in strong magnetic fields. Therefore, for weak, $\lesssim 1\%$, oscillations one can write:
\begin{equation}
\frac{R_{\mathrm{osc}}}{R_{\mathrm{bg}}} = -\frac{\sigma_{\mathrm{osc}}}{\sigma_{\mathrm{bg}}}\,,
\label{Rsigma}
\end{equation}
where $\sigma_{\mathrm{osc}}=\sigma_{\mathrm{osc},\alpha}+\sigma_{\mathrm{osc},\beta}$ and $\sigma_{\mathrm{bg}}$ is the non-oscillating part of the interlayer conductivity. Eq.\,(\ref{Rsigma}) can be brought to a form suitable for the LK analysis by noting that
\begin{equation}
\frac{\sigma_{\mathrm{osc}}}{\sigma_{\mathrm{bg}}} =
\frac{\sigma_{\mathrm{osc},\alpha}}{\sigma_{\mathrm{bg},\alpha}}\cdot
\frac{\sigma_{\mathrm{bg},\alpha}}{\sigma_{\mathrm{bg}}} +
\frac{\sigma_{\mathrm{osc},\beta}}{\sigma_{\mathrm{bg},\beta}}\cdot
\frac{\sigma_{\mathrm{bg},\beta}}{\sigma_{\mathrm{bg}}}\,.
\label{sigma}
\end{equation}
Thus, we can fit the resistance oscillations in the framework of the LK formalism, provided the relative conductivity contributions $\sigma_{\mathrm{bg},\alpha}/\sigma_{\mathrm{bg}}$ and $\sigma_{\mathrm{bg},\beta}/\sigma_{\mathrm{bg}}$ are known.

As the MB $\beta$ oscillations obviously involve all the carriers on the Fermi surface, the corresponding background is simply the total conductivity: $\sigma_{\mathrm{bg},\beta} = \sigma_{\mathrm{bg}} \propto 1/R_{\mathrm{bg}}$. According to the experimental results, \cite{helm09} the high-field magnetoresistance is approximately quadratic in $B$. While this field dependence may look counterintuitive at first glance (the charge transport along the magnetic field direction is often believed to be unaffected by the field), it is a direct consequence of the symmetry properties of the body-centered tetragonal (b.c.t.) structure of NCCO. It can be shown \cite{berg03} that, if a cylindrical Fermi surface of a strongly anisotropic layered metal with a b.c.t. lattice is centered in the corner of the first Brillouin zone $(\pi/a,\pi/a,k_z)$, the magnetic field direction perpendicular to the layers satisfies Yamaji's magic angle condition: \cite{yama89}
all the cyclotron orbits on the Fermi surface have the same area. In this case the interlayer conductivity should decrease \cite{kart92,grig10} $\propto 1/B^2$, in agreement with the experiment.

The conduction channel $\sigma_{\alpha}$ is associated with small pockets centered at points $(\pm \pi/2a,\pm\pi/2a,k_z)$ in the Brillouin zone. For these parts of the Fermi surface the field directed perpendicular to the layers is away from a Yamaji angle. According to the standard theory, \cite{abri88b} $\sigma_{\mathrm{bg},\alpha}(B)$ saturates at a level close to the zero field value, so for the calculations it was assumed to be field-independent, $\sigma_{\mathrm{bg},\alpha}(B)=\sigma_{\mathrm{bg},\alpha}(0)$. The relative contribution of the small pockets to the total interlayer conductivity, $\sigma_{\mathrm{bg},\alpha}(0)/\sigma_{\mathrm{bg}}(0)$, has been calculated using the classical transport Boltzmann equation \cite{abri88b} and the tight-binding dispersion relation
\begin{equation}
\epsilon\left(\mathbf{k}_{\|},k_z,\varphi\right) = \epsilon_{\|}\left(\mathbf{k}_{\|}\right) - 2t_{\perp}(\varphi)\cos(dk_z)\,,
\label{disp}
\end{equation}
where $\mathbf{k}_{\|}$ and $k_z$ are the in-plane and out-of-plane components of the electron wave vector, $d\approx 0.6$\,nm is the distance between adjacent CuO$_2$ layers, $\varphi$ the azimuthal angle of $\mathbf{k}$ in the $k_xk_y$-plane, and $t_{\perp} \ll \epsilon_{\|}$. The in-plane dispersion $\epsilon_{\|}\left(\mathbf{k}_{\|}\right)$ was taken from literature. \cite{lin05} For the simplest case of a $\varphi$-independent $t_{\perp}$, we have estimated $\sigma_{\mathrm{bg},\alpha}(0)/\sigma_{\mathrm{bg}}(0)=0.39$. A more realistic, $\varphi$-dependent interlayer transfer term complying with the b.c.t. lattice symmetry, $t_{\perp}(\varphi)= t_{\perp,0}(\sin2\varphi+0.3\sin6\varphi-0.3\sin10\varphi)$, was obtained from the analysis of the angle-dependent magnetoresistance oscillations, \cite{helm13} which will be published separately. Substituting this $t_{\perp}(\varphi)$ in Eq.~(\ref{disp}) results in the relative contribution of the $\alpha$ pockets $\sigma_{\mathrm{bg},\alpha}(0)/\sigma_{\mathrm{bg}}(0)=0.42$, i.e. only slightly different from the simplest estimate given above.

\end{appendix}

\bibliographystyle{apsrev}

%\bibliography{HTSCnew}

\begin{thebibliography}{55}
\bibitem[{com({\natexlab{a}})}]{comment_Dingle}
\bibinfo{note}{It should be noted that our fitting assumes uniform scattering
  over the whole Fermi surface, so that the Dingle temperature is the same for
  the $\alpha$ and $\beta$ orbits. We believe that this is a reasonable
  assumption for the present low-temperature ($\simeq 2$\,K) data, since the
  $\mathbf{k}$-dependent scattering channel has been shown to vanish at
  temperatures $4-5$\,K for another b.c.t. cuprate superconductor
  Tl$_2$Ba$_2$CuO$_{6+\delta}$ \cite{huss06}. Moreover, the good fits achieved
  simultaneously for the fast and slow oscillations provide additional strong
  support for this assumption.}
\bibitem[{com({\natexlab{b}})}]{comment_Hall}
\bibinfo{note}{A weak curvature of $\rho_{xy}(B)$ at fields between 20 and
  50\,T can be attributed to an unavoidable finite (typically $\leq 0.3\%$)
  distribution width of Ce concentration through the sample volume. Note that
  at fields $> 50$\,T the signal comes to a linear negative slope extrapolating
  to zero (cf. straight dashed line in Fig.\,\ref{Hall_B}), indicating that no
  MB occurs in the major part of the sample even at the highest fields.}
\end{thebibliography}

\end{document}